\documentclass[%
 aip,
rsi,%
 amsmath,amssymb,
 preprint,%
]{revtex4-1}
\pdfoutput=1
\usepackage{graphicx}
\usepackage{dcolumn}
\usepackage{bm}
\usepackage{amsmath}
\usepackage{amssymb}
\usepackage{color}
\usepackage{soul}

\usepackage{listings}
\usepackage{amsmath}
\usepackage{graphicx} 
\usepackage{epstopdf}
\usepackage{amssymb} 
\usepackage{gensymb} 

\usepackage{color}

\newlength\figureheight
\newlength\figurewidth

\newcommand{\ve}[1]{\boldsymbol{#1}}  
\newcommand{\md}{\mathrm{d}} 

\graphicspath{{./images/}}

\begin{document}

\title{Curvature controlled defect dynamics in topological active nematics}

\author{Francesco Alaimo}
\affiliation{Institute of Scientific Computing, Technische Universit\"at Dresden, 01062 Dresden, Germany}
\affiliation{Dresden Center for Computational Materials Science (DCMS), 01062 Dresden, Germany}

\author{Christian K\"ohler}
\affiliation{Institute of Scientific Computing, Technische Universit\"at Dresden, 01062 Dresden, Germany}

\author{Axel Voigt}
\email[Corresponding author: ]{axel.voigt@tu-dresden.de.}

\affiliation{Institute of Scientific Computing, Technische Universit\"at Dresden, 01062 Dresden, Germany}
\affiliation{Dresden Center for Computational Materials Science (DCMS), 01062 Dresden, Germany}
\affiliation{Center of Systems Biology Dresden (CSBD), Pfotenhauerstr. 108, 01307 Dresden, Germany}

\date{\today}

\begin{abstract}
We study the spatiotemporal patterns that emerge when an active nematic film is topologically constraint. These topological constraints allow to control the non-equilibrium dynamics of the active system. We consider ellipsoidal shapes for which the resulting defects are 1/2 disclinations and analyze the relation between their location and dynamics and local geometric properties of the ellipsoid. We highlight two dynamic modes: a tunable periodic state that oscillates between two defect configurations on a spherical shape and a tunable rotating state for oblate spheroids. We further demonstrate the relation between defects and high Gaussian curvature and umbilical points and point out limits for a coarse-grained description of defects as self-propelled particles.

\end{abstract}

\maketitle
 
 Active systems are characterized by constant input of energy, which is converted by autonomous constituents into directed motion, leading to spatiotemporal 
patterns. These phenomena range from the macroscale, e.g. flocks of birds \cite{Cavagnaetal_PNAS_2010} or schools of fish \cite{Hemelrijketal_BE_2005} to the microscale, e.g. bacterial colonies
\cite{Wensinketal_PNAS_2012}, migrating tissue cells \cite{Szaboetal_PRE_2006} or active nematic films \cite{Giomietal_PRL_2013}. If such systems are confined on curved surfaces, topological constraints strongly
influence the emerging spatiotemporal patterns. Using these topological constraints to guide collective cell behavior might be a key in morphogenesis and active nematic films 
on surfaces have been proposed as a promising road to engineer synthetic materials that mimic living organisms \cite{Keberetal_Science_2014}. However, the complex 
dynamics of such topological active systems remains wildly unexplored. As in passive systems the mathematical Poincar\'e-Hopf theorem forces topological defects to be present in the nematic film. On a sphere this leads to an equilibrium defect configuration with four +1/2 disclinations arranged as a tetrahedron 
\cite{Lubenskietal_JP_1992,Shinetal_PRL_2008,Skacejetal_PRL_2008}, see Figure 1 
The disclinations repel each other and this arrangement maximizes their distance. In active systems unbalanced stresses drive this configuration out of equilibrium. But in contrast to planar active 
nematics with continuous creation and annihilation of defects \cite{Sanchezetal_Nature_2012,DeChampetal_NatureM_2015}  the creation of additional defect pairs can be suppressed on curved surfaces, which is demonstrated in  \cite{Keberetal_Science_2014} for an active nematic film of microtubules and molecular motors, encapsulated within a spherical lipid vesicle. This provides an unique way to study the 
dynamics of the four defects in a controlled manner and led to the discovery of a tunable periodic state that 
oscillates between the tetrahedral and a planar defect configuration. We confirm this finding by computer simulations, see Figure 1. 

Within a coarse-grained model +1/2 disclinations in planar active nematic films can be effectively described by self-propelled particles with a velocity proportional to the activity \cite{Giomietal_PRL_2013}. In \cite{Keberetal_Science_2014} this relation is extended to spherical nematics. Four self-propelled particles on a sphere also oscillate between 
the planar and tetrahedral configuration. Both descriptions can be quantitatively linked to each other, but also differences can be pointed out, which become more evident for more general surfaces. For non-constant Gaussian curvature constraints local geometric properties influence the position of the defects and thus can be used to control defect dynamics. We are concerned with a systematic investigation of the impact of such constraints on the emergence of complex patterns and oscillations.

\section*{Results}

For active systems in flat geometries various theoretical descriptions have been proposed, see e.g. \cite{Marchettietal_RMP_2013,Bechingeretal_RMP_2016}. One of the most studied
approaches are Vicsek-like models \cite{Visceketal_PRL_1995}. They consider particles, which travel at a constant speed to represent self-propulsion, whose direction 
changes according to interaction rules which comprise explicit alignment and noise. In contrast to equilibrium systems long-range order emerges for two dimensional systems with low noise. We consider an extension of these models which includes excluded volume \cite{Tailleuretal_PRL_2008,Henkesetal_PRE_2011,Filyetal_PRL_2012,Bialkeetal_PRL_2012} and classify systems by the head-tail symmetry of their particles in polar or nematic. For active polar particles these models have been formulated on a sphere
\cite{Sknepneketal_PRE_2015} and on ellipsoidal surfaces \cite{Ehrigetal_arXiv_2016}. In these situations a robust rotating-band structure around the waist, with two +1 
defects at the poles is found on a sphere. On an ellipsoid the location of the defects is linked to local geometric properties, similar to vortices in surface fluids \cite{Turneretal_RMP_2010,Nitschkeetal_JFM_2012,Reutheretal_MMS_2015,Nitschkeetal_arXiv_2016}. The defects are related to the Gaussian curvature and to the umbilical points of the surface (see Materials and Methods for a geometric description). For spheroidal ellipsoids there are two umbilical points, which locate the two +1 defects. This configuration is more stable for prolate spheroids, where the umbilical points are at the points of maximal Gaussian curvature at the poles and less stable for oblate spheroids, where the umbilical points and the maximum in Gaussian curvature are separated. As in the spherical case a rotating-band structure is formed, with possible sub-bands which counter rotate depending on the initial condition. New dynamical features are found for non-spherical ellipsoids. They have four umbilical points. For lower velocities the defects encircle pairs of umbilical points and for larger velocities the defects are found at the high Gaussian curvature regions between each pair of umbilical points. With this richness in dynamics found for active polar particles on non-constant Gaussian curvature surfaces, we expect similar behavior for active nematic particles and ask up to which complexity of the geometry the dynamics of the four 1/2 disclinations can be effectively described by self-propelled polar particles. 

\begin{figure}
\centering
\includegraphics[width=.7\linewidth]{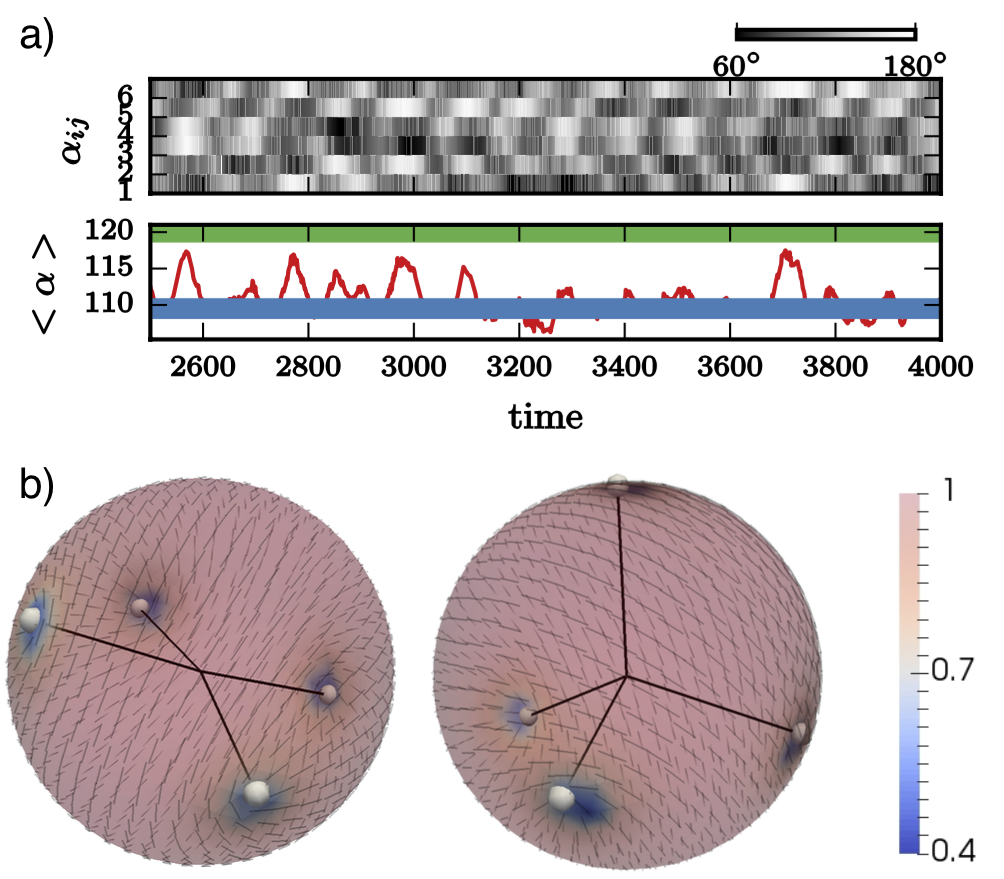}
\caption{{Defect oscillations: } a) Top: Kymograph showing the time evolution of the angles $\alpha_{ij}$, which denote the angle between the radii from the center of the sphere 
to each of the defect pairs. Bottom: Oscillation of the average angle $< \alpha >$. The blue and the green line correspond to the planar ($< \alpha > = 120^\circ$) 
and tetrahedral ($< \alpha > = 109,5^\circ$) defect configuration. b) Snapshots showing the planar and tetrahedral defect configuration within a simulation of 1.000
particles (the four 1/2 disclinations are highlighted, the director field is shown - black lines - and the color coding corresponds to the nematic order parameter $P$, with minima in the 
four defects). The results are in excellent agreement with the experimental results in \cite{Keberetal_Science_2014}. A video is provided in the SI.}
\label{fig1}
\end{figure}

To answer these questions, we first analyze the spherical case in more detail. In addition to the oscillation between the planar and  tetrahedral defect configuration on a spherical vesicles and a tunable frequency by the activity and self-propulsion velocity we also track the positions of the defects. Computing the power spectrum from the time series for the average angle $< \alpha > = \frac{1}{6} \sum_{i < j} \alpha_{ij}$ we obtain the frequency for the oscillations, which linearly depend on the activity. The same results, but with a small offset and a different slope are obtained for the coarse-grained description by self-propelled particles, see Figure 2. 
As a consequence for each activity in the nematic film a self-propulsion velocity can be determined in the coarse-grained description, which resamples the frequency of the planar-tetrahedral defect oscillation. Differences between both descriptions are found if we compare the trajectories of the defects and self-propelled particles. Within the considered time interval the 1/2 disclinations are locally confined, each defect only covers part of the vesicle. This is in contrast to the trajectories of the self-propelled particles, which rotate within a band structure leaving parts of the vesicle uncovered, see Figure 2. 
The experimental defect trajectories in \cite{Keberetal_Science_2014} differ from both descriptions, they are global, covering the whole vesicle. The discrepancy might be a consequence of the considered short-range interactions in the model for the active nematic film.

\begin{figure}
\centering
\includegraphics[width=.7\linewidth]{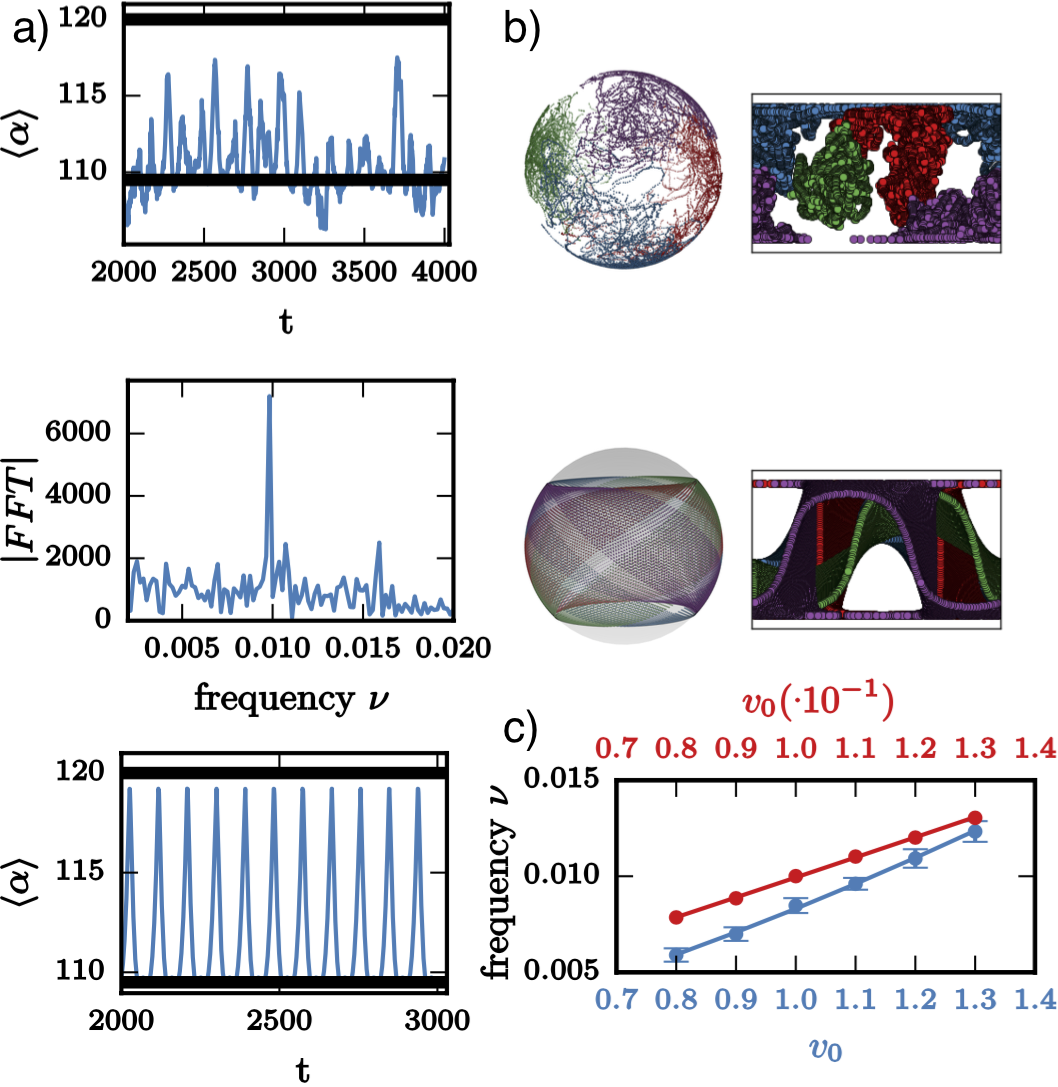}
\caption{{ Model comparison: } a) Top: Oscillation of the average angle $< \alpha >$ from Fig. 1, Middle: The power spectrum of $< \alpha >$ obtained by using the Fast Fourier Transform (FFT), the peak is associated with the planar-tetrahedral oscillations, Bottom: Oscillation of the average angle $<\alpha>$ for four self-propelled particles. b) Top: Trajectories of the four 1/2 disclinations, each color corresponds to one defect, shown on the sphere and using the Gall-Peters projection, Bottom: same as Top but for the four self-propelled particles.
c) Frequency for the planar-tetrahedral oscillation corresponding to the peak in the power spectrum as a function of the activity for various realizations (blue curve). The trajectories of the four self-propelled particles show a perfect planar-tetrahedral oscillation, the frequency is obtained as the distance between consecutive maxima and shown as a function of the self-propulsion velocity (red curve).}
\label{fig2}
\end{figure}

We next consider spheroidal ellipsoids. They are characterized by the aspect ratio $a/c$ and $a = b$, with $a$, $b$ and $c$ the length of the major axis. Due to the symmetry all geometric properties can be characterized with respect to the polar axis. As the geometry is topologically equivalent to a sphere we expect for passive systems again a minimal energy configuration with four 1/2 disclinations. They still try to maximize their distance, but are now also influenced by local geometric properties. The 1/2 disclinations tend to accumulate in regions of high Gaussian curvature \cite{Batesetal_SM_2009,Serra_LC_2016}. Computer simulations for thin nematic shells have shown that for prolate ellipsoids pairs of defects are located at opposite ends close to the poles. The defects in each pair arrange at opposite sides of the surface and tend to align perpendicular to the pair at the other pole \cite{Batesetal_SM_2009}. As the distance between the defects is no longer maximized, the geometric effect seems to dominate the repulsion in this case. For oblate ellipsoids the 1/2 disclinations are found near the waist, where the Gaussian curvature is largest. Again two pairs of defects are found, one on each side. They repel each other and are mutually perpendicular to the other pair, leading to an alternating ring of 1/2 disclinations, one above and one below the waist. This behavior seems to be independent of the film thickness \cite{Batesetal_SM_2009}, we have confirmed this behavior by our surface model without activity.

\begin{figure}
\centering
\includegraphics[width=.7\linewidth]{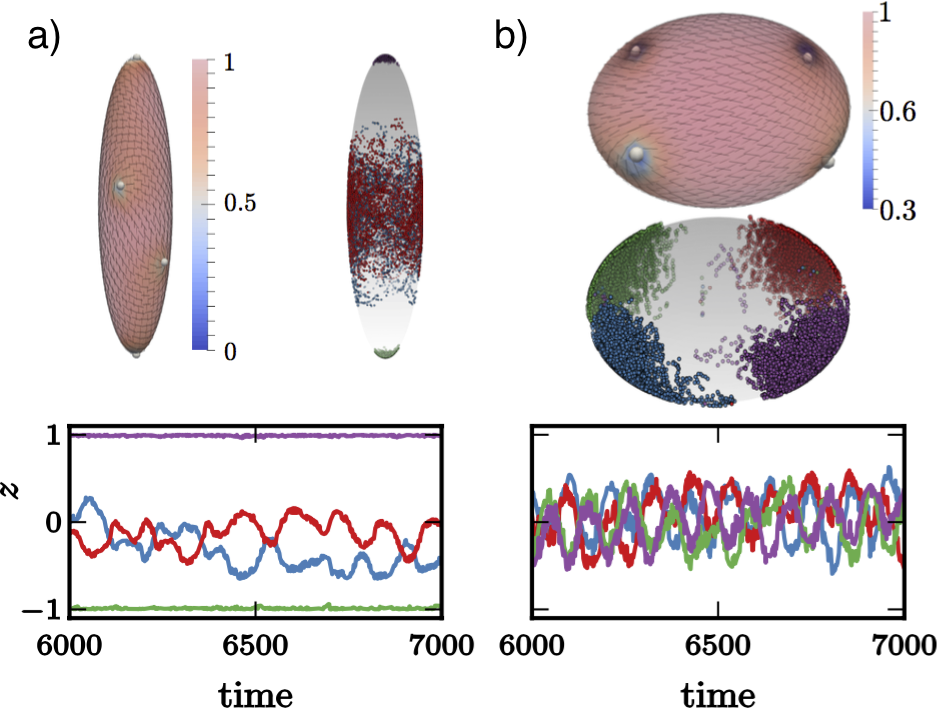}
\caption{{ Defect localization on spheroids: } a) Snapshot showing the defect configuration within a simulation of 1.000 particles on a prolate spheroid with $a/c = 0.25$ 
(the four 1/2 disclinations are highlighted, the director field - black lines - is shown and the color coding corresponds to the nematic order parameter $P$, 
with minima in the defects). In addition the trajectories of the four 1/2 disclinations are shown (each color corresponds to one defect). The height $h_i$ 
for each defect with respect to the waist is also shown as a function of time. b) same as a) for a oblate spheroid with $a/c = 2 $. The oscillations of the four defects have the same frequency and alternate with respect to each other. Videos for a) and b) are provided in the SI.}
\label{fig3}
\end{figure}

For active systems we observe again oscillatory behavior, see Figure 3. 
For prolate spheroids ($a/c < 1$) only two 1/2 disclinations are located at the poles, whereas the other two oscillate around the waist. The oscillations are very noisy and can not be tuned by the activity. Even if the distance between the two 1/2 disclinations at the waist is not optimal the average distance between all four defects is larger than in the passive case. While the 1/2 disclinations are still attracted by the high curvature regions at the poles, the active forces push one of the defects away leading to the observed metastable configuration. Within a transition zone ($ a/c \approx 1$) we observe similar behavior as in the spherical case ($a/c = 1$) without any defect localization. The behavior changes for oblate spheroids ($a/c > 1$), where all four 1/2 disclinations are along the waist, maintaining a maximal distance to each other. This behavior is similar to the passive system. However, the defects now oscillate between both sides. The frequency of the alternating oscillations above and below the waist can be extracted for various activities. However, a clear functional dependency on the activity could not be found. If the aspect ratio is further increased the situation changes to pairs of 1/2 disclinations which rotate around the umbilical points at the poles. The defects are no longer located at positions of maximal Gaussian curvature. The high curvature value at the waist creates a distortion of the nematic film, which can be seen from the nematic order parameter. It somehow serves as a barrier for the 1/2 disclinations preventing them from crossing the waist. The rotation is a consequence of the activity and the unfavorable short distance with respect to each other. The frequency of the rotation depends on the activity and can be tuned, see Figure 4. 
Also the transition to this rotating state depends on the strength of the activity. As stronger the activity as longer it is possible for the defects to cross the barrier at the waist. A tendency to locate the defects away from the high Gaussian curvature waist can also be seen for the passive case.

\begin{figure}
\centering
\includegraphics[width=.7\linewidth]{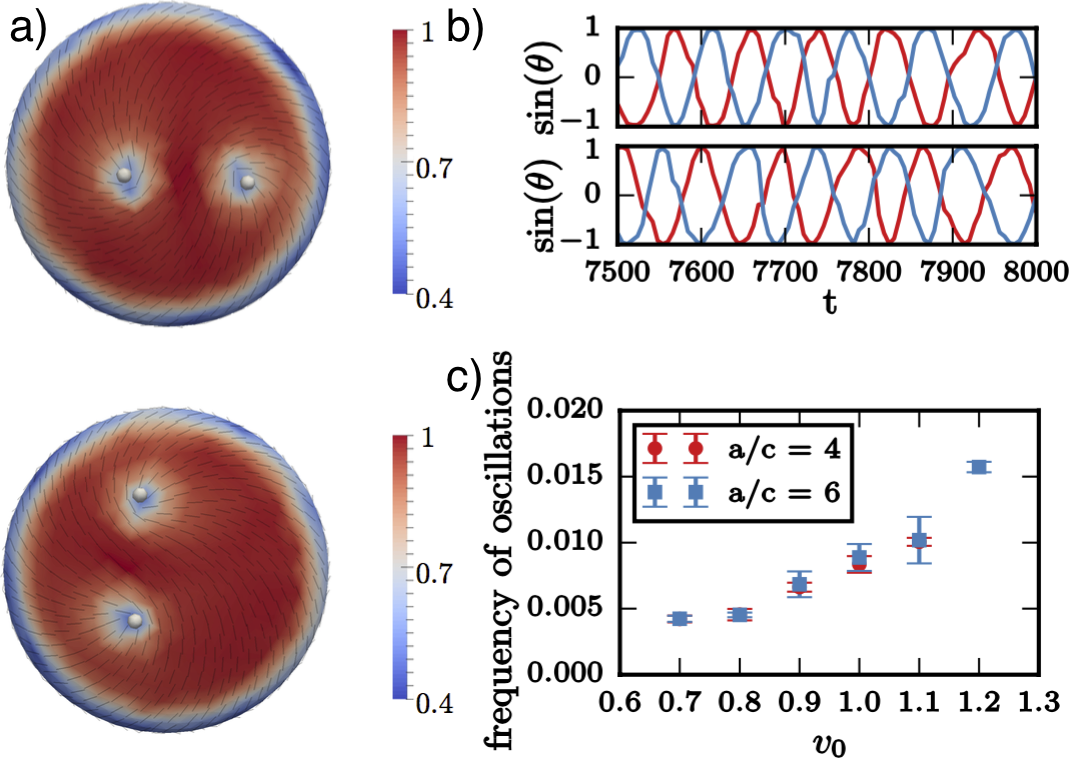}
\caption{{ Defect rotations: } a) Snapshots from above and below showing the defect configurtion within a simulation with 1.000 particles on an oblate spheroid with $a/c = 6$ (the four 1/2 disclinations are highlighted, the director field - black lines - is shown and the color coding corresponds to the nematic order parameter $P$, 
with minima in the defects). b) Oscillations of the angle measuring the rotation around the umbilical points (top and bottom) and c) frequency of the oscillation as a function of activity for two different aspect ratios. A video for case $a/c = 6$ is provided in the SI.}
\label{fig4}
\end{figure}

The four different regimes are shown in Figure 5. 
using the order parameter 
\[
\eta = \frac{1}{4 N c} \sum_{i = 1}^4 \sum_{t = ts}^{te} |h_i(t)|,
\] 
with $N$ the number of particles, $[ts,te]$ an appropriate time interval and $h_i(t)$ the height of the defect $i$ along the polar axis with respect to the waist at time $t$. We have $\eta = 1$ if all defects are at the poles, $\eta = 0$ if they are at the waist and $\eta = 0.5$ if they are homogeneously distributed along the polar axis.  

Within the coarse-grained description by self-propelled polar particles, using the corresponding self-propulsion velocity according to Figure 2, 
we obtain a qualitatively different behavior. Within the considered parameter regime, the values for $\eta$ are independent of the self-propulsion velocity. For aspect ratios $a/c < 0.5$ the particles rotate on closed trajectories, well separated from each other at approximately equal distance along the polar axis. The transition zone with sphere-like behavior is more extended than for the nematic defects. For $0.5 < a/c < 2$ a band structure is formed around the waist, which shrinks with increasing aspect ratio. For $a/c > 2$ all particles are positioned at the waist, rotating in one direction and maintaining their distance. The regime with pairwise rotating defects around the umbilical points could not be found within the coarse-grained model (A more detailed description is given in the SI). 

\begin{figure}
\centering
\includegraphics[width=.7\linewidth]{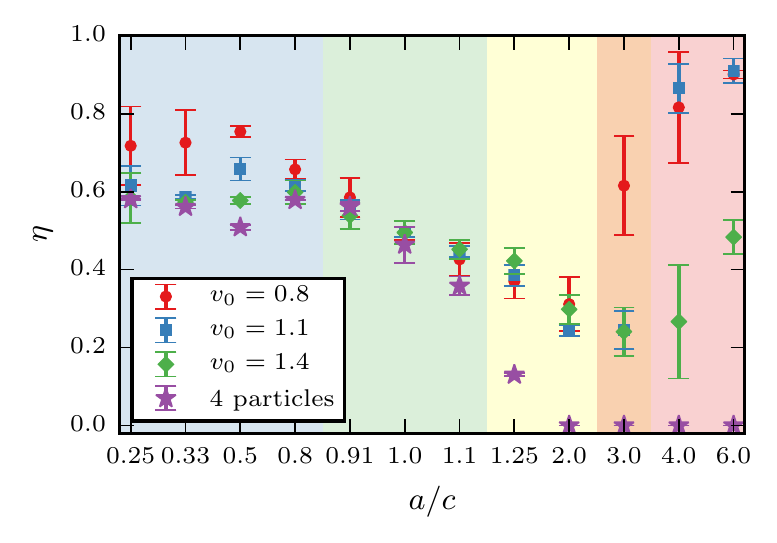} 
\caption{{ Phase diagram: } Phase diagram for patterns and oscillations on spheroidal ellipsoids for 1/2 disclinations and self-propelled particles. The results for the coarse-grained description by self-propelled particles are independent of the activity in the corresponding regime to the considered velocities $v_0$. From left to right we have (blue) the situation for prolate shapes with location of two defects at the poles, leading to $\eta > 0.5$, (green) spherical like shapes with no clear location of the defects, leading to $\eta \approx 0.5$, (yellow) oblate shapes with location of the defects along the waist, leading to $\eta < 0.5$, for larger $a/c$ we obtain a phase transition towards the rotating state, with the defects located around the poles, leading to $\eta > 0.5$. The transition towards this state depends on the activity.}
\label{fig5}
\end{figure}

Non-spherical ellipsoids, which are characterized by $a \neq b$, $a \neq c$ and $b \neq c$, have four umbilical points. They are either prolate-like or oblate-like but in any case have two distinct points of maximal Gaussian curvature. We thus analyze the distance of the four 1/2 disclinations with respect to the umbilical points and the points of maximal Gaussian curvature using the average geodesic distances $<DDU>$ and $<DDG>$, respectively. Figure 6, 
which is inspired by \cite{Ehrigetal_arXiv_2016} shows the distances as a function of the aspect ratios $a/b$ and $a/c$. Spheroids are also included, the first column shows the previous results for oblate and the diagonal for prolate geometries. Each row in between thus corresponds to a transition from oblate-like to prolate-like geometries. In most cases the 1/2 disclinations are closer to the high Gaussian curvature points than to the umbilical points, with the only exception for oblate-like ellipsoids   with a large aspect ration $a/c \geq 4$. This leads to the conclusion that 1/2 disclinations tend to be attracted by points of high Gaussian curvature. 

\begin{figure}
\includegraphics[width=.7\linewidth]{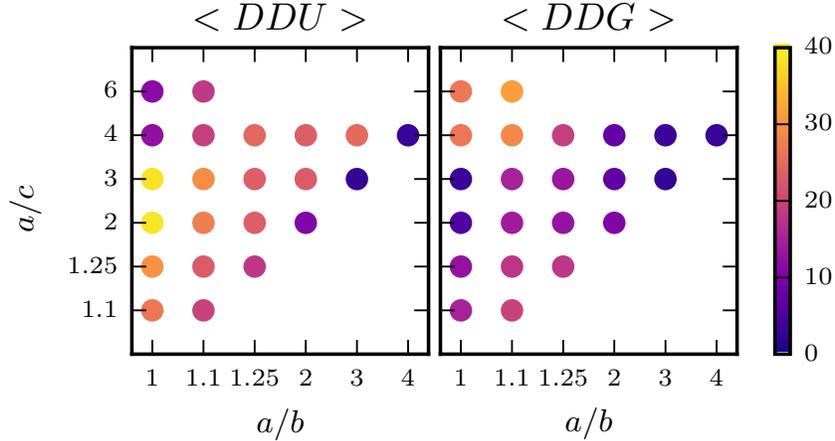}
\caption{{ Relation to geometric properties: } Average geodesic distance of 1/2 disclinations to the umbilical points $<DDU>$ (left) and to the points of maximal Gaussian curvature $<DDG>$ (right) for non-spheroidal and spheroidal (first column - oblate and diagonal - prolate) ellipsoids of different aspect ratio. Only for the extreme case of a/c = 4,6 and a/b = 1.1 the disclinations are closer to the umbilical points. Also in these cases a rotating state as in Figure \ref{fig4} can be observed, which however is not as regular, see SI. In all other situations the disclinations are closer to the points of maximal Gaussian curvature. A video for case $a/b = 1.1$ and $a/c = 6$ is provided in the SI.}
\label{fig6}
\end{figure}

\section*{Discussions}

In \cite{Keberetal_Science_2014} it was shown that in a confined active system, a dense suspension of microtubules and molecular motors on the surface of a spherical lipid vesicle, cyclic oscillations between defect configurations can be observed. They result from topological constraints and the coupling between velocity fields and defect-defect interactions. These findings may push forward the design of systems that harness the ability of nanoscale active matter to transform chemical energy into mechanical work. On non-spherical surfaces defects are known to be strongly influenced by local geometric properties. The induced geometric interaction can lead to locating of defects, which is established for vortices in surface fluids \cite{Turneretal_RMP_2010,Nitschkeetal_JFM_2012,Nitschkeetal_arXiv_2016} and vortices, sources and sinks in polar systems \cite{Nestleretal_arXiv_2016,Ehrigetal_arXiv_2016}. For strong variations in geometric properties it has even be found computationally that lower energy minima in passive systems can be formed by creating additional defects \cite{Reutheretal_MMS_2015,Nestleretal_arXiv_2016} for surface fluids and surface polar particles, respectively. Our work extends the understanding of the delicate relations between topology, geometry and defect dynamics on non-spherical shapes for the system considered in  \cite{Keberetal_Science_2014}. We are concerned with ellipsoidal surfaces and identify crucial geometric features which influence collective motion patterns in active nematic films. We have shown that 1/2 disclinations are related to both, maxima in the Gaussian curvature and umbilical points of the surface. On prolate spheroids maxima in Gaussian curvature and umbilical points coincide, they are located at the two poles and attract the 1/2 disclinations. However, the repulsive defect-defect interaction allows only two of the defects to be located at the poles, the other two try to maximize their distance and are located around the waist, where they oscillate. Spherical like shapes lead to similar behavior as observed on a sphere, with no distinguished location of the defects and an oscillation between a tetrahedral and planar defect configuration. For oblate spheroids all 1/2 disclinations are located at the waist, the region of high Gaussian curvature. They again maximize their distance and oscillate. With increasing aspect ratio $a/c$ the situation changes. The defects can no longer cross the waist, where the high Gaussian curvature leads to a distortion of the nematic order. As a consequence pairs of 1/2 disclinations rotate around the umbilical points. The frequency of the rotation depends on the activity and can be tuned. This found rotating state is an other step towards a controllable transformation of chemical into mechanical energy in nanoscale active matter and asks for experimental validation. The results for non-spheroidal ellipsoids confirm these findings, even if the separation of the different states is not as distinct as in Figure 5. 
A smooth transition of the dynamics between prolate-like and oblate-like shapes is identified in Figure, 
with a clear tendency of the 1/2 disclinations to locate at points of maximal Gaussian curvature. Only for extrem values of $a/c$ and almost spheroidal shapes the situation changes and the rotating state around the umbilical points could be identified.   

We further demonstrate that the proposed coarse-grained description of 1/2 disclinations in active nematic matter by self-propelled particles fails if geometric properties come into play. Already on spherical shapes the trajectories of the defects and the self-propelled particles differ significantly and on spheroidal ellipsoids both descriptions not even qualitatively agree.    

In summary we explored the complex interaction of topology, geometry and defect dynamics in nematic films on ellipsoidal surfaces and demonstrated how topological constraints and geometric properties can be used to control the collective behavior in nanoscale active matter. The non-linear coupling between non-constant Gaussian curvature and defect-defect interactions leads to tunable spatiotemporal patterns. Among these findings is a stable rotating state on strongly oblate-like ellipsoids, which suggests an other pathway towards a controllable generation of mechanical work in nanoscale active matter. The richness of physics observed in our work will further increase if the underlying shape is deformable. First experimental results of such an interplay between activity-driven defect motion and deformability of the vesicle are already shown in \cite{Keberetal_Science_2014} and discussed in \cite{Yeomans_NM_2015}. However, for theoretical descriptions of these phenomena new methods will be required.     

\section*{Materials and Methods}

We consider a more general approach than the Viscek-like models confined on a sphere or an ellipsoids in \cite{Sknepneketal_PRE_2015,Ehrigetal_arXiv_2016}. 

\subsection{Equations of motion}
We consider $N$ active particles of mass $m_i=1$, which are constrained to move on a surface algebraically described by $g(\mathbf{q}) = 0$, with particle positions $\mathbf{q} = \left(\mathbf{q}_1, \ldots, \mathbf{q}_N\right)$. Newton's equations of motion (EOM) with holomonic constraint $g(\mathbf{q})$ read:
 \begin{equation}\label{EQ::dynamic}
 \begin{array}{ll}
   \displaystyle \frac{\md}{\md t} \mathbf{q} = \mathbf{v}, \quad   
   \displaystyle \frac{\md}{\md t} \mathbf{v} = \mathbf{F} - G\left(\mathbf{q}\right)^T \boldsymbol{\lambda} ,  \quad   
   \displaystyle g\left(\mathbf{q}\right) = \mathbf{0}
 \end{array}
 \end{equation}
with forces $\mathbf{F} = (\mathbf{F}_1, \ldots, \mathbf{F}_N)$ and velocities $\mathbf{v} = (\mathbf{v}_1, \ldots, \mathbf{v}_N)$. $\boldsymbol{\lambda} = (\lambda_1, \ldots, \lambda_N)$ are the Lagrange multipliers and $G(\mathbf{q}) = \nabla_{\mathbf{q}} g(\mathbf{q})$ is the Jacobian of $g(\mathbf{q})$.
 The force $\mathbf{F}_i$ can be written as:
 \begin{equation}
  \mathbf{F}_i = - \gamma \mathbf{v}_i+ \sum_{j=1}^N\mathbf{F}_{ij} + \mathbf{F}_i^{ac}  
 \end{equation}
 where $\gamma$ is the translational friction coefficient, $\mathbf{F}_i^{ac}$ the active force acting on particle $i$ and 
 $\mathbf{F}_{ij}$ the pair-interaction force between particle $i$ and particle $j$. Additionally every particle has an internal degree of freedom, its orientation $\mathbf{n}_i$. Denoting by $\ve{\omega}_i$ the angular velocity we have the following 
 EOM for the orientational dynamics:
 \begin{equation}\label{EQ::orientational_dynamic}
 \begin{array}{ll}
   \displaystyle \frac{\md}{\md t} \mathbf{n}_i = \ve{\omega}_i\times \mathbf{n}_i, \quad 
   \displaystyle \frac{\md}{\md t} \ve{\omega}_i = -\gamma_a \ve{\omega}_i + \mathbf{T}_i\left(\mathbf{q},\mathbf{n}\right) 
 \end{array}
 \end{equation}
 where $\gamma_a$ is the rotational friction coefficient and $\mathbf{T}_i(\mathbf{q}, \mathbf{n})$ is the torque acting on particle $i$, with $\mathbf{n} = (\mathbf{n}_1, \ldots, \mathbf{n}_N)$. 
 Depending on the specific 
 form for the active force $\mathbf{F}_i^{ac}$, the pair-interaction force $\mathbf{F}_{ij}$, the torque $\mathbf{T}_i(\mathbf{q}, \mathbf{n})$ and the 
 holomonic constraint $g(\mathbf{q})$ we will be able to describe polar and nematic active systems on various surfaces.
 
 \subsection{Active polar particles}
 For active polar particles on a 
 sphere of radius $R$ we would specify \cite{Sknepneketal_PRE_2015} $\mathbf{F}_i^{ac} = v_0 \mathbf{n}_i$ with a constant self-propulsion velocity $v_0$, 
 $\mathbf{F}_{ij} = k (2 \sigma - q^g_{ij}) \frac{\mathbf{q}_i - \mathbf{q}_j}{q_{ij}}$ for $q^g_{ij} < 2 \sigma$ and 
 $\mathbf{F}_{ij} = 0$ otherwise, a short-range repulsion between spheres of radius 
 $\sigma$, with elastic constant $k$, euclidian distance $q_{ij} = |\mathbf{q}_i - \mathbf{q}_j|$ and geodesic distance $q^g_{ij} = |\mathbf{q}_i - \mathbf{q}_j|_g$.
 Parallel orientations between neighbouring particles are favored and therefore we use the aligning torque 
 $\mathbf{T}_i(\mathbf{q}, \mathbf{n}) = - J \sum_{j \in U(i)} \left ( \mathbf{n}_i \times \mathbf{n}_j \right )$, 
 with $J > 0$ the strength and $U(i)$ the first shell of neighbors of particle $i$, identified as all the particles
 within a cutoff radius of $2.4 \sigma$ from $\mathbf{r}_i$.
 The holomonic constraint for a sphere of radius $R$ reads $g(\mathbf{q}_i) = q_{i,1} ^2 + q_{i,2}^2 + q_{i,3}^2 - R^2$, with 
 $\mathbf{q}_i = (q_{i,1},q_{i,2}, q_{i,3}) \in {\cal R}^3$. 
 This approach can be used to reproduce the results in \cite{Sknepneketal_PRE_2015} in which the overdamped limit, 
 the euclidian distance instead of the geodesic distance and an additional noise term are considered. 
 
 \subsection{Active nematic particles}
 To describe active nematic particles we use the tensor order
 parameter $Q_{\alpha \beta}^j = \left ( n_{\alpha}^j n_{\beta}^j - \delta_{\alpha \beta}/3 \right) $, where the upper index corresponds to the particles 
 and the lower indices represent the componenets $x, y, z$.
 The active force does not distinguish 'head from tail' and it thus has the form:
 \begin{equation}
 \mathbf{F}_i^{ac} = - v_0 \sum_{j \in U(i)}  \mathbf{Q}^j \frac{ \mathbf{q}_i - \mathbf{q}_j }{q_{ij}^2}.
 \end{equation}
 The torque reflects the fact that both parallel and anti-parallel configurations are favored. It has the form:
 \begin{equation}
 \mathbf{T}_i = J \sum_{j \in U(i)} \left ( \left ( \mathbf{n}_i \cdot \mathbf{n}_j \right ) \left (\mathbf{n}_i \times \mathbf{n}_j \right ) \right ).
 \end{equation}
 The pair-interaction force $\mathbf{F}_{ij}$ and the holomonic constraint $g(\mathbf{q})$ are the same as in the active polar particles case.
 The simulation parameters for this case are $(J, k, \sigma, \gamma, \gamma_a, v_0) = (10, 3, 2, 0.1, 2.5, 1.1)$ unless otherwise specified.
 
 \subsection{Coarse-grained defect description}
 In the coarse-grained defects description by active polar particles \cite{Giomietal_PRL_2013} and \cite{Keberetal_Science_2014} 
 the elastic energy between defects is $E \sim \log(q_{ij}^g)$, where $q_{ij}^g$ is the geodesic distance between the defects.
 The pair-interaction force is therefore $ \mathbf{F}_{ij} = \frac{k}{q_{ij}^g} \frac{\mathbf{q}_i - \mathbf{q}_j}{q_{ij}}$,
 which is no longer short-ranged. 
 Defects align anti-parallel to each other  and the restoring torque strength is \cite{Keberetal_Science_2014} $T_i = J \sum_{j \in U(i)} \cot( \frac{\theta_{ij}}{2})$, where
 $\theta_{ij}$ is the angle between $\mathbf{n}_i$ and $\mathbf{n}_j$. The vector form for the torque can be written in terms of the orientations as:
 \begin{equation}
  \mathbf{T}_i = J \sum_{j \in U(i)} \left ( 1 + \mathbf{n}_i \cdot \mathbf{n}_j \right ) \frac{\mathbf{n}_i \times \mathbf{n}_j}{|\mathbf{n}_i \times \mathbf{n}_j|^2}
 \end{equation}
 Finally the defects are treated as self-propelled particles and the active force is $\mathbf{F}_i^{ac} = v_0 \mathbf{n}_i$.
 The simulation parameters for this case are $(J, k, \gamma, \gamma_a, v_0) = (3, 4, 0.1, 2.5, 0.11)$ unless otherwise specified.

 \subsection{Geometric properties}
 Besides a sphere we consider two classes of ellipsoidal surfaces: (i) spheroidal and (ii) non-spheroidal. 
 These ellipsoids are characterized by their major axis $a$, $b$ and $c$ and have non-constant Gaussian curvature
 \begin{equation}
  K = \frac{a^2 b^6 c^6}{\left (c^4 b^4 + c^4 (a^2 - b^2) y^2 + b^4 (a^2 - c^2) z^2 \right)^2}.
 \end{equation}
 For spheroidal ellipsoids two of these values are equal. The algebraic description reads 
 $g(\mathbf{q}_i) = \frac{q_{i,1}^2}{a^2} + \frac{q_{i,2}^2}{b^2} + \frac{q_{i,3}^2}{c^2} - 1 = 0$. 
 An umbilic point is a point where the maximum and minimum curvatures coincide. 
 At an umbilical point, the surface is "locally spherical". These points are found at
 \begin{equation}
  \left(\pm a \sqrt{\frac{a^2 - b^2}{a^2 - c^2}}, 0 , \pm c \sqrt{\frac{b^2 - c^2}{a^2 - c^2}} \right)^T
 \end{equation}
 In figure 7 
 we show three different ellipsoids, where umbilical points are highlighted and the color coding corresponds to the 
 Gaussian curvature $K$.
 
 \begin{figure}
 \centering
 \includegraphics{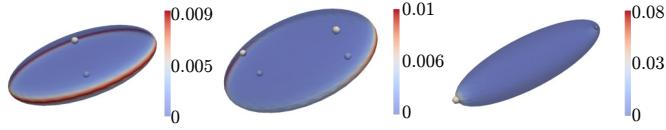}
 \caption{{ Geometric features: } Example of an ellipsoid with major axis a) $ a/b = 1$ and $ a/c = 4$, b) $a/b = 1.25$ and $a/c = 4$ and c) $a/b = 1$ and $a/c = 0.25$ . 
 Umbilical points are shown as points and the Gaussian curvature $K$ is color coded.}
 \label{fig7}
 \end{figure}
 
 \subsection{Numerical methods}
 Eqs. \ref{EQ::dynamic} have been numerically solved using RATTLE discretization \cite{HamDyn}.
 The equation for the orientational dynamic eqs. \ref{EQ::orientational_dynamic} have been first solved unconstrained with
 the torque $\mathbf{T_i}$ projected on to the normal plane of the surface at point $\mathbf{r}_i$.
 Afterwards the orientation $\mathbf{n}_i$ has been projected on to the tangent plane of the surface at point $\mathbf{r}_i$ 
 and the angular velocity $\ve{\omega}_i$ takes the direction of the normal to the surface at point $\mathbf{r}_i$.
 
 We fix the number of particles $N = 1000 $ and the volume fraction $\phi \simeq 1 $. The surface area for the sphere is equal to $A = 4 \pi R^2$,
 with $R = 31.6$. The ellipsoid parameters $a,b,c$ have been chosen such that the surface area is equivalent to the surface area of the sphere
 and the aspect ratio is respected.
 
 The nematic order parameter is defined as 
 \[
 P_i = \frac{1}{\sum_j w_{ij}} \sum_j \frac{w_{ij}}{2} 
 ( 3 \ve{n}_i \cdot \ve{n}_j - 1 )
 \] 
where the sum is over the nearest neighbors and $w_{ij} = q_{ij}^{-1}$. 
 
 Defects are calculated as the local center of mass for regions where the local order parameter $P_i$ is smaller then $0.45$ 
 (some corrections were required for regions of high Gaussian curvature, due to strong distortion of the director field).
 
 The simulation code is implemented in C++, using the GeographicLib library \cite{karney2011} for the calculation of the geodesic distances. However,
 for non-spheroidal ellipsoids the euclidian distance has been used. This approximation can be justified by the short-range interactions.
 Data have been analyzed using Python, Ovito \cite{ovito} and Paraview.
  \begin{acknowledgments}
This work was funded by the European Union (ERDF) and the Free State of Saxony via the ESF project 100231947 (Young Investigators Group Computer Simulation for Materials Design - CoSiMa). We used computing resources provided by JSC within project HDR06.
 \end{acknowledgments}

\textbf{Author contributions} F.A. and C.K. derived the model and conducted the simulations, F.A. and A.V. analyzed the results.  All authors contributed to a critical discussion of the results and participated in writing the manuscript, which A.V. finalized.

\bibliographystyle{unsrt} 
\bibliography{bibliography}

\end{document}